\documentclass[journal=aamickt,manuscript=article]{achemso}
\setkeys{acs}{usetitle = true}
\setkeys{acs}{keywords = true}
\usepackage{chemformula} 
\usepackage[T1]{fontenc} 

\SectionNumbersOn
\usepackage{notes2bib}

\usepackage{wrapfig}
\usepackage{epsfig}
\usepackage{amsthm}
\usepackage{hyperref}
\usepackage{amsmath}
\usepackage{cleveref}
\usepackage{CJKutf8}
\usepackage{latexsym}
\usepackage{amstext}
\usepackage{amsfonts}
\usepackage{dsfont}
\usepackage{color}
\usepackage{amssymb}
\usepackage{relsize}
\usepackage{amsxtra}
\usepackage{verbatim}
\usepackage{hyperref}
\usepackage{graphics}
\usepackage{graphicx}
\usepackage{multirow}
\usepackage{ifpdf}
\usepackage{lineno}
\usepackage{bbding}

\usepackage{caption}
\usepackage[labelfont=bf]{caption}
\usepackage[labelsep=space]{caption}
\author{Zirui Zhao}
\affiliation{Institute of Applied Physics and Materials Engineering, University of Macau, Avenida da Universidade, Taipa, Macao SAR 999078, China.}
\author{Hai-Feng Li}
\affiliation{Institute of Applied Physics and Materials Engineering, University of Macau, Avenida da Universidade, Taipa, Macao SAR 999078, China.}
\email{haifengli@um.edu.mo}
\phone{+853 8822 4035}
\fax{+853 8822 2426}

\title{Investigating Material Interface Diffusion Phenomena through Graph Neural Networks in Applied Materials}


\begin{document}

\begin{tocentry}

\includegraphics[width=\textwidth]{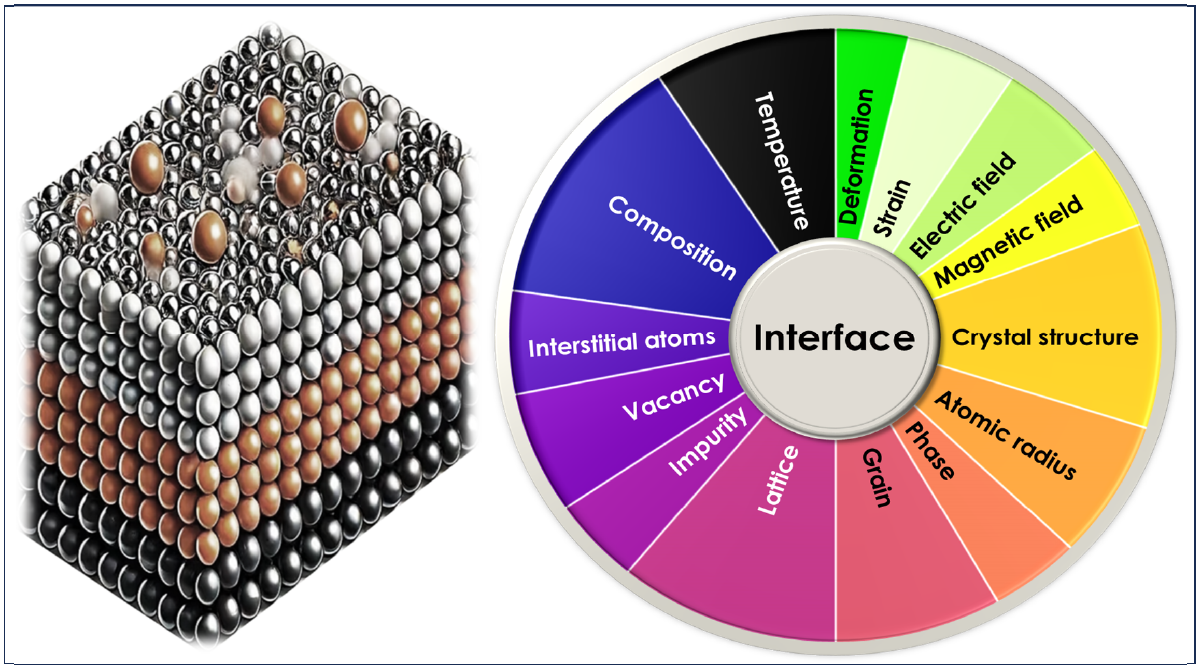} \\
\medskip
\noindent{\textbf{Caption of Graphical Abstract:} By analyzing experimental and simulated data, GNNs accurately forecast diffusion coefficients, rates, and pathways, enhancing material design and performance.}
\bigskip

\end{tocentry}

\begin{abstract}
Understanding and predicting interface diffusion phenomena in materials is crucial for various industrial applications, including semiconductor manufacturing, battery technology, and catalysis. In this study, we propose a novel approach utilizing Graph Neural Networks (GNNs) to investigate and model material interface diffusion. We begin by collecting experimental and simulated data on diffusion coefficients, concentration gradients, and other relevant parameters from diverse material systems. The data are preprocessed, and key features influencing interface diffusion are extracted. Subsequently, we construct a GNN model tailored to the diffusion problem, with a graph representation capturing the atomic structure of materials. The model architecture includes multiple graph convolutional layers for feature aggregation and update, as well as optional graph attention layers to capture complex relationships between atoms. We train and validate the GNN model using the preprocessed data, achieving accurate predictions of diffusion coefficients, diffusion rates, concentration profiles, and potential diffusion pathways. Our approach offers insights into the underlying mechanisms of interface diffusion and provides a valuable tool for optimizing material design and engineering. Additionally, our method offers possible strategies to solve the longstanding problems related to materials interface diffusion.
\end{abstract}
\section{Introduction}

Materials science is a pivotal discipline driving modern technological advancements, focusing on the composition, structure, properties, and relationships of materials. Within materials engineering, interface diffusion profoundly influences mechanical properties, conductivity, and thermal stability, impacting various practical applications.\cite{gupta1998interface, goodenough2010challenges} However, traditional research methods, relying on experimental techniques and numerical simulations, often struggle with complex multiphase interface problems due to their time-consuming and labor-intensive nature.\cite{ward1952diffusion} Despite significant advancements in materials science and engineering, many unresolved challenges persist in the study of interfacial diffusion. The complexity of interfacial phenomena, driven by diverse material properties and environmental conditions, continues to elude comprehensive understanding. Moreover, the heterogeneous nature of materials and the multi-scale interactions involved further complicate accurate prediction and control of diffusion processes at interfaces. 

To address these challenges, it is essential to identify specific unresolved issues related to material interface diffusion in various material systems, as detailed in Table.~\ref{table:unresolved_issues}. These issues are critical in the context of materials science and engineering, especially in applications involving semiconductors, electronic devices, solid-state batteries, and thin-film technologies.\cite{ZHOU2023101248, https://doi.org/10.1002/ejic.202300382, XIAO2024151111, WANG202479} For instance, the diffusion at the interface between silicon (Si) and silicon dioxide (SiO\textsubscript{2}) significantly affects the electrical properties in semiconductor manufacturing, posing challenges in enhancing device performance.\cite{fukatsu2004effect} Similarly, the electromigration phenomenon due to copper (Cu) interconnect diffusion in silicon (Si) impacts circuit stability and resistance, requiring advanced strategies to mitigate these effects.\cite{takeyama1996properties} In the case of germanium-doped silicon (Ge-Si), understanding the interface diffusion is essential for managing crystal defects and optimizing electronic properties.\cite{silvestri2006diffusion} Additionally, multilayer nitride films on silicon need careful examination of diffusion impacts to maintain film stress and structural stability.\cite{chang1982diffusion} High-power electronic devices benefit from controlled silicon and silicon carbide (SiC) interface diffusion,\cite{komninou1999gold} while the dielectric properties and adhesion of aluminum oxide (Al\textsubscript{2}O\textsubscript{3}) thin films depend on minimizing interface diffusion with silicon.\cite{werner2011electronic} Furthermore, the performance of high-frequency devices can be compromised by the diffusion at the silicon and gallium nitride (GaN) interface.\cite{lin2000improved} The contact resistance and reliability issues arising from gold (Au) diffusion in silicon,\cite{slezak2000surface} along with the electrical properties of nickel silicide (NiSi) affected by nickel (Ni) diffusion, highlight the need for precise interface management.\cite{chang1982diffusion} Finally, optimizing the performance of optoelectronic devices involves addressing silicon and indium phosphide (InP) interface diffusion challenges.\cite{shapira1984interdiffusion} These examples underscore the importance of advanced research and innovative solutions in managing material interface diffusion to enhance the performance and reliability of various technological applications.

In recent years, the rapid advancement of artificial intelligence, particularly in the realm of machine learning, offers promising avenues for materials science. Graph Neural Networks (GNNs), adept at handling non-euclidean data such as graph-structured data, have emerged as a novel approach with distinct advantages.\cite{gori2005new} GNNs not only adeptly capture complex relationships within material microstructures but also efficiently predict and optimize material properties.\cite{bruna2013spectral} In this context, this paper investigates the application of GNN technology to study diffusion phenomena at material interfaces. Through the development of an efficient GNN model, the aim is to achieve more accurate predictions within shorter timeframes, thereby furnishing novel insights and methodologies for materials science research.\cite{wang2023capacity,verma2018graph}

This study endeavors to utilize GNN technology to address diffusion challenges at material interfaces, with specific objectives including the development of a GNN-based model capable of accurately capturing complex material microstructure relationships, simulating and predicting interface diffusion behavior, and validating the model's effectiveness and reliability. Additionally, the study aims to explore key factors influencing the diffusion process and analyze their mechanisms in diverse material systems, furnishing theoretical foundations for material design and optimization. To achieve these objectives, the study employs various research methods, including data acquisition and preprocessing to collect and clean experimental and simulation data on interface diffusion, model construction and training to build a predictive model based on GNN, and result analysis and validation. Most importantly, from the trained GNN model, we extracted potential strategies for shedding light on presently unsolved problems related to materials interface diffusion.

\section{Graph neural networks and applications}

\subsection{Fundamentals and advantages of graph neural networks}

GNNs represent a sophisticated class of neural network models specifically engineered to operate on graph-structured data. These networks offer a robust framework for encoding intricate relationships between entities within high-dimensional data matrices.\cite{griggs2020unified} The foundational strength of GNNs lies in their adeptness at capturing and understanding complex relationships inherent in graph-structured data. 

Leveraging a mathematical formulation, GNNs facilitate the seamless propagation of information between neighboring nodes, thereby enabling iterative updates to node embeddings through a message-passing mechanism defined as:
\begin{equation}
h_v^{(k)} = \text{Aggregate}\left(\{h_u^{(k-1)} : u \in \mathcal{N}(v)\}\right),
\end{equation}
where $h_v^{(k)}$ represents the embedding of node $v$ at iteration $k$, $\mathcal{N}(v)$ denotes the neighborhood of node $v$, and $\text{Aggregate}$ denotes the aggregation function. Moreover, GNNs incorporate pooling operations, which facilitate the aggregation of information from multiple nodes to generate comprehensive representations of entire graphs. This pooling operation can be mathematically expressed as:
\begin{equation}
h_G = \text{Pool}\left(\{h_v^{(K)} : v \in V\}\right),
\end{equation}
where $h_G$ represents the graph-level embedding, $V$ denotes the set of all nodes in the graph, and $\text{Pool}$ denotes the pooling function.

The ability of GNNs to model and learn from high-dimensional structured data, coupled with their computational efficiency, positions them as versatile tools applicable across various domains. Notably, GNNs find extensive utility in applications such as molecular structure prediction and recommender systems, where their capacity to grasp intricate relationships within data proves invaluable.\cite{griggs2020unified} For instance, in molecular structure prediction, GNNs can model the interactions between atoms and predict molecular properties with high accuracy. In recommender systems, GNNs can capture user-item interactions to provide personalized recommendations.

\subsection{Case studies of graph neural networks in material property prediction}

GNNs demonstrate remarkable proficiency in discerning intricate patterns within molecular structures and accurately forecasting properties such as energy, stability, and reactivity. This surpasses the capabilities of traditional data-driven algorithms, heralding a new era in materials prediction. For instance, recent advancements in materials science have witnessed the emergence of GNNs as formidable instruments for predicting material properties based on atomic structure. These neural network models offer a paradigm shift by representing molecules as graphs, with atoms serving as nodes and bonds as edges, thereby enabling a nuanced description of materials' microscopic constituents and structural attributes.\cite{chen2023md, sathana2022prediction}

To better understand the advantages of GNNS, it is essential to compare them with traditional simulation methods. Traditional methods, such as molecular dynamics (MD) and density functional theory (DFT), often require extensive computational resources and time, particularly for complex systems with large numbers of atoms and intricate interactions.\cite{liu2004calculation} These methods also depend heavily on accurate potential energy functions and may face challenges in capturing long-time scale phenomena due to their inherent limitations.\cite{patel2019computing} In contrast, machine learning algorithms can leverage large datasets to learn and predict complex diffusion behaviors with high accuracy and efficiency. By training on existing experimental and simulation data, machine learning models can generalize to new conditions and materials without the need for exhaustive recalculations. This ability to generalize makes machine learning particularly powerful for predicting diffusion properties across a wide range of materials and interface conditions.\cite{allers2020machine} Additionally, machine learning models can incorporate a broader set of features, including atomic environments, local chemistry, and external conditions, to provide more comprehensive predictions. These models can also be updated and improved continuously as new data becomes available, enhancing their predictive power over time. Moreover, machine learning techniques can uncover hidden patterns and relationships in diffusion processes that may not be apparent through traditional methods, offering novel insights and guiding experimental design more effectively. Overall, the integration of machine learning algorithms into material interface diffusion studies presents a transformative approach, enabling faster, more accurate, and more insightful analysis compared to conventional simulation techniques.

Beyond diffusion studies, the application spectrum of GNNs in materials science extends beyond mere prediction; they play pivotal roles in catalyst design by modeling catalyst atomic structures and interactions with reactant molecules, thus predicting catalytic activity and selectivity. Similarly, in battery material prediction and design, GNNs leverage atomic structure and crystallographic information to predict crucial properties such as capacity, cycle stability, and rate capability, thereby streamlining research efforts in this domain. Furthermore, recent studies underscore the superiority of GNNs over traditional machine learning models in predicting material properties such as bandgaps in semiconductor crystals, thereby offering valuable insights into structure-property relationships and accelerating materials discovery and design endeavors.\cite{zhang2023graph}

\section{Research methods}
\subsection{Data acquisition and preprocessing}

The first step in our research involves acquiring and preprocessing the data necessary for training and validating our GNN model. The data are collected from both experimental and simulation sources, focusing on various material systems and their interface diffusion characteristics.

\textbf{Data Collection:} Experimental data are collected from literature, databases, and experimental reports detailing the diffusion coefficients, concentration gradients, and other relevant parameters of different materials. Additionally, we utilize molecular dynamics (MD) simulations and finite element analysis (FEA) to generate supplementary data, providing detailed atomic-level insights into the diffusion processes.\cite{jiangwei2002microstructure, svoboda2011diffusion, yang2020comprehensive, janssen1973diffusion, chambers1987silicide, peng2005study, divinski2004ag, erdelyi2002interface, burger1987structure, li2014molecular, ogawa1995generalized, shen2001electronic, lee1997prediction, lee2014two, whitlow1991diffusion, adalsteinsson2003transport, tavoosi2017kirkendall, luthra1991mixed, trumble1991thermodynamics, watanabe1994electron, langer1975theory, griscom1985diffusion} This comprehensive collection ensures a robust dataset for subsequent analysis.

\textbf{Data Cleaning:} We remove any incomplete, redundant, or inconsistent data entries to ensure the quality and reliability of the dataset. Furthermore, the data formats and units are standardized to maintain uniformity across the dataset. This step is crucial for minimizing errors and enhancing the accuracy of our model.

\textbf{Feature Extraction:} Key features that influence interface diffusion, such as atomic radius, electronegativity, lattice parameters, and bonding characteristics, are extracted. 

To determine the importance of various descriptors in predicting material interface diffusion, we employed Principal Component Analysis (PCA). PCA is a dimensionality reduction technique that transforms the original set of features into a new set of uncorrelated components (principal components), which capture the most variance in the data. In our analysis, we applied PCA to the dataset containing the various descriptors related to material properties and diffusion parameters. By examining the loadings of the original features on these principal components, we were able to assess the contribution of each descriptor to the overall variance in the diffusion phenomena. The percentage values shown in Figure 1 represent the relative importance of each descriptor, as derived from its contribution to the most significant principal components.
This approach allowed us to objectively identify which descriptors were most influential in predicting the diffusion behavior, providing a clear understanding of the factors driving the observed phenomena. The use of PCA thus ensured that the most critical features were highlighted, facilitating a more accurate and interpretable model.
\color{black}

\begin{figure*} [!t]
\centering \includegraphics[width=0.82\textwidth]{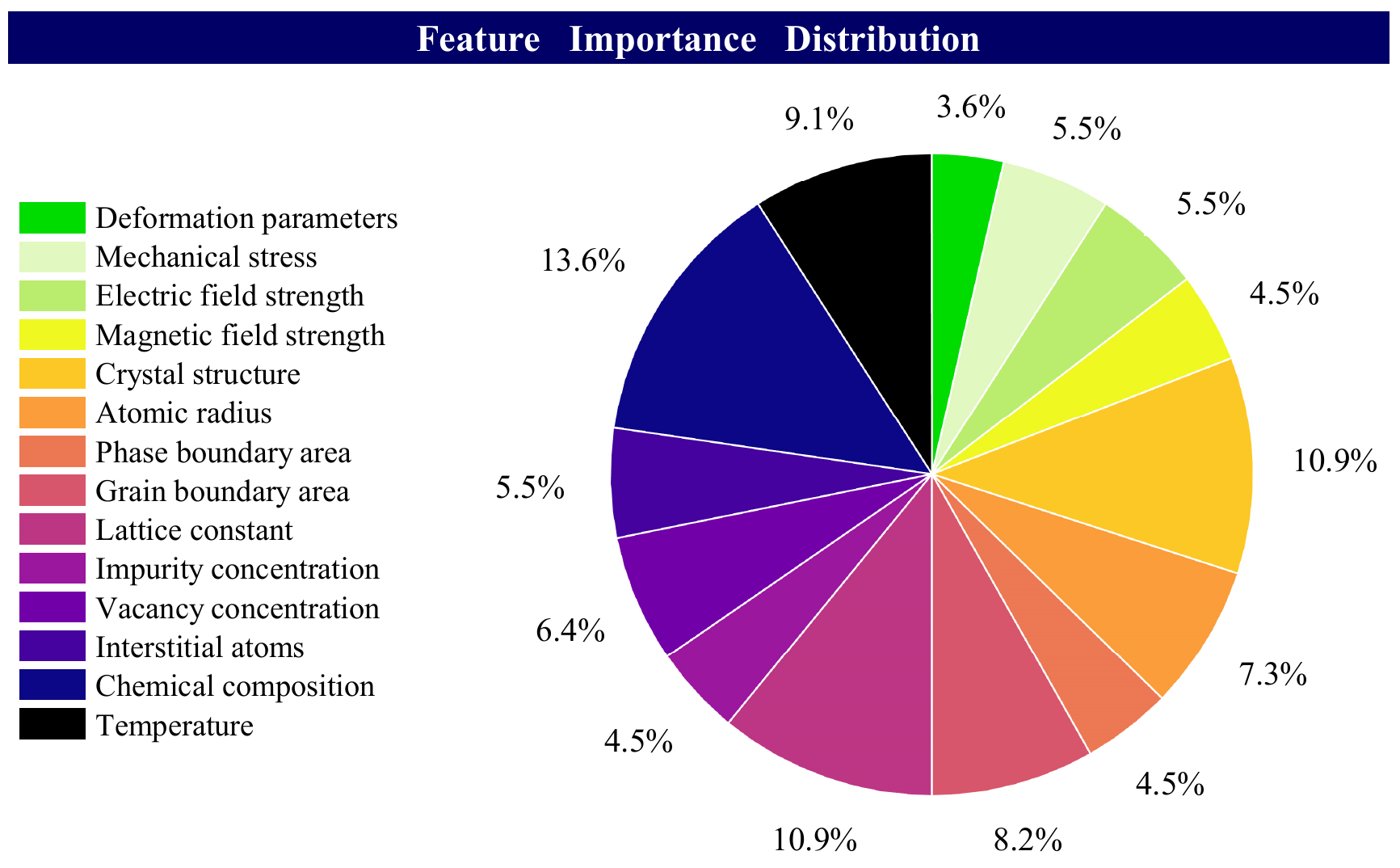}
\caption{Schematic representation of the distribution of feature importance, including all input parameters, and their respective influence percentages on the output.}
\label{Feature}
\end{figure*}

Figure~\ref{Feature} illustrates the impact of each feature on the output parameters after training. It is evident that the chemical components have the greatest impact on the output parameters, with a 13.6$\%$ influence. This is followed by crystal structure and lattice constants, both of which have a 10.9$\%$ impact on the output parameters. The importance of these features in the model's predictions indicates their key role in determining material properties and provides an important reference for subsequent studies.

\subsection{Feature engineering and input data preparation}

To predict material interface diffusion efficiently using machine learning, we need to handle the identified influential factors appropriately as input features for our model. Key features such as atomic radius and lattice constant are included as input features, using normalized or standardized values as necessary. The vacancy concentration of materials under specific conditions, derived from experimental data or simulations, is also included. Types and concentrations of interstitial atoms are considered as input features. The temperature at which the material is studied is included, normalized or standardized as necessary. Chemical composition is represented using one-hot encoding or embedding vectors, and the types and concentrations of impurities are included as input features. Grain boundary area or grain size and phase boundary area or phase boundary density are also included. The type of crystal structure is encoded as a numerical feature (embedding vectors). Values of mechanical stress experienced by the material and deformation levels or strain are included as input features. The strength and direction of any applied electric field and magnetic field are also considered as input features.\cite{mackiewicz1993principal} These features ensure a comprehensive representation of the factors influencing diffusion.

\begin{figure*} [!t]
\centering \includegraphics[width=0.82\textwidth]{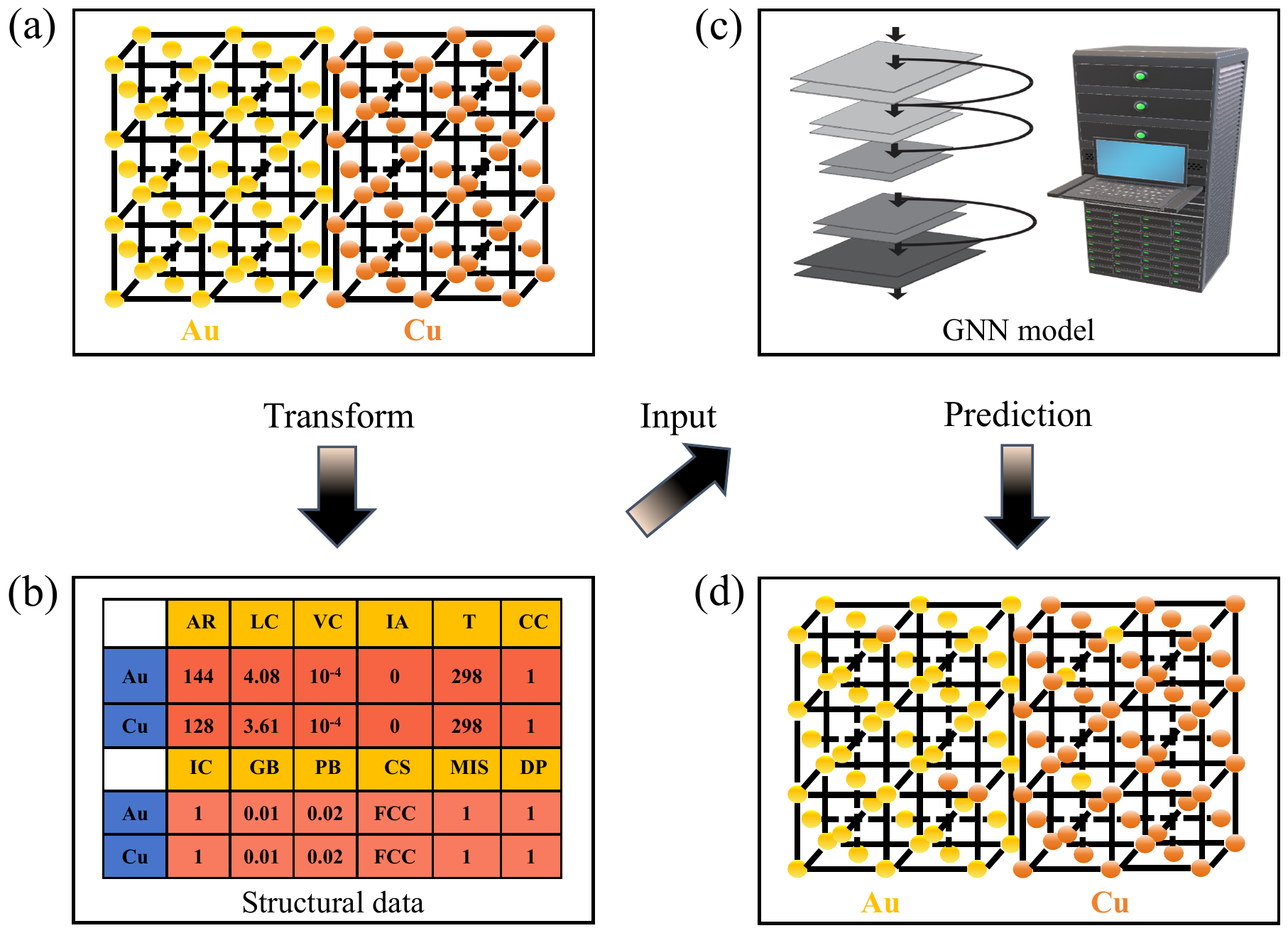}
\caption{Workflow of this study, illustrating the interfacing materials of two phases represented as a matrix and input into the GNN model for predicting interface diffusion phenomena. In this figure, (a) shows the actual crystal structure at the contact interface, using Au and Cu as examples. (b) Abstracts this structure into a matrix form for computational modeling. (c) Demonstrates the matrix as input into the model for training purposes. (d) Displays the predicted outcomes derived from the trained model, Where Au, Cu atoms left their original lattice positions to complete diffusion.}
\label{Workflow}
\end{figure*}

Figure~\ref{Workflow} illustrates the detailed workflow of our study. Initially, the researchers represent the interfacing materials of the two phases in a matrix form, capturing the essential structural and compositional information. This matrix representation serves as the input for the corresponding GNN model. The model processes this input to predict the interface diffusion phenomena, leveraging its capability to handle high-dimensional graph-structured data and to capture complex interactions between the materials at the interface.

In this study, we employed a comprehensive dataset to train and validate our GNN model aimed at predicting material interface diffusion characteristics. The input parameters for the model were carefully selected to ensure a thorough representation of the factors influencing diffusion processes. The key input parameters included atomic radius, electronegativity, diffusion coefficient, temperature, concentration gradient, grain boundary energy, defect density, crystal structure, and interface orientation. These parameters were chosen based on their significant impact on diffusion behavior, as identified in prior research. This careful selection ensures the model's robustness and accuracy. To facilitate the model's learning process, these input parameters were preprocessed and normalized. The GNN model was designed to capture the intricate relationships between these parameters and the resultant diffusion characteristics. The output parameters of the model encompassed the primary metrics used to evaluate diffusion, such as diffusion coefficient, activation energy for diffusion, and diffusivity as a function of temperature and concentration. These outputs were critical for assessing the model's predictive performance and for comparing its predictions with experimental and theoretical values. This comprehensive approach ensures a thorough evaluation of the model's capabilities.

\subsection{Construction of graph neural network models}

Once the data is preprocessed, the next step is constructing the GNN model tailored to our specific problem of interface diffusion in materials.

\textbf{Input Layer:} Features of each atom in the material, such as atomic radius, atom type (represented by embedding vectors), and local environment features, are included as node features. Features of each bond in the material, such as bond length, bond type, and bond energy, are included as edge features.

\textbf{Graph Convolution Layers:} Next, multiple graph convolution layers (GCNs) are utilized to aggregate features from nodes and their neighbors, capturing the dependencies between nodes. The graph convolution operation is given by:
\begin{equation}
\mathbf{h}_v^{(k+1)} = \sigma \left( \sum_{u \in \mathcal{N}(v)} \frac{1}{c_{vu}} \mathbf{W}^{(k)} \mathbf{h}_u^{(k)} + \mathbf{W}_0^{(k)} \mathbf{h}_v^{(k)} \right), 
\end{equation}
where \( \mathbf{h}_v^{(k)} \) is the feature representation of node \( v \) at layer \( k \), \( \mathcal{N}(v) \) is the set of neighboring nodes of \( v \), \( c_{vu} \) is a normalization constant, \( \mathbf{W}^{(k)} \) and \( \mathbf{W}_0^{(k)} \) are learnable weight matrices, and \( \sigma \) is a non-linear activation function.

There are several key components involved in building a GNN model for predicting material properties. The core of the GNN is the graph convolutional layer, which aggregates information from neighbouring nodes in the graph. The update rule for the graph convolution layer can be defined as:
\begin{equation}
h_i^{(l+1)} = \sigma \left( \sum_{j \in N(i)} \frac{1}{c_{ij}} W^{(l)} h_j^{(l)} \right). 
\end{equation}
In the equation, \( h_i^{(l)} \) represents node \( i \)'s representation at layer \( l \), \( N(i) \) is the set of neighboring nodes of node \( i \), \( c_{ij} \) is a normalization constant for the edge between nodes \( i \) and \( j \), \( W^{(l)} \) is the weight matrix at layer \( l \), and \( \sigma \) is a non-linear activation function. 

\textbf{Aggregation Layer:} Following the GCNs, global pooling aggregates node-level features into a graph-level feature representation, generating a fixed-length vector representing the entire material structure. Node representations are aggregated into graph-level representations using readout functions. Common readout functions include sum, mean, or maximum pooling, expressed as the following equation: 
\begin{equation}
h_{\text{graph}} = \text{READOUT} \left( \{ h_i^{(L)} \}_{i \in V} \right). 
\end{equation}
In the equation, \( h_{\text{graph}} \) is the graph-level representation, \( \{ h_i^{(L)} \}_{i \in V} \) are the final layer node representations, and \( \text{READOUT} \) is the readout function. 

\textbf{Fully Connected Layers:} Subsequently, several fully connected layers process the aggregated feature vector, capturing higher-level feature combinations. The graph-layer representation predicts the desired material properties through multiple fully connected layers, as shown in the equation below: 
\begin{equation}
\hat{y} = \text{FC} \left( h_{\text{graph}} \right). 
\end{equation}
In the equation, \( \hat{y} \) is the predicted material property and \( \text{FC} \) is a fully connected layer. The predicted material property \( \hat{y} \) is compared with the truth property \( y \) using a suitable loss function, such as mean squared error (MSE), as shown in the equation below: 
\begin{equation}
\mathcal{L} = \text{Loss} \left( \hat{y}, y \right). 
\end{equation}
Consequently, through backpropagation and gradient descent algorithms, GNN models are continuously trained to minimize the loss function.

\textbf{Output Layer:} The output layer is set based on the task type. For regression tasks, diffusion coefficients, diffusion rates, etc., are output using a linear activation function. For classification tasks, diffusion pathways or other classification results are output using a softmax activation function.

\textbf{Graph Representation:} Finally, the material system is represented as a graph \( G = (V, E) \), where \( V \) represents the nodes (atoms) and \( E \) represents the edges (bonds or interactions between atoms). The initial features of the nodes and edges are defined based on the extracted features from the data preprocessing step.

\begin{figure*} [!t]
\centering \includegraphics[width=0.82\textwidth]{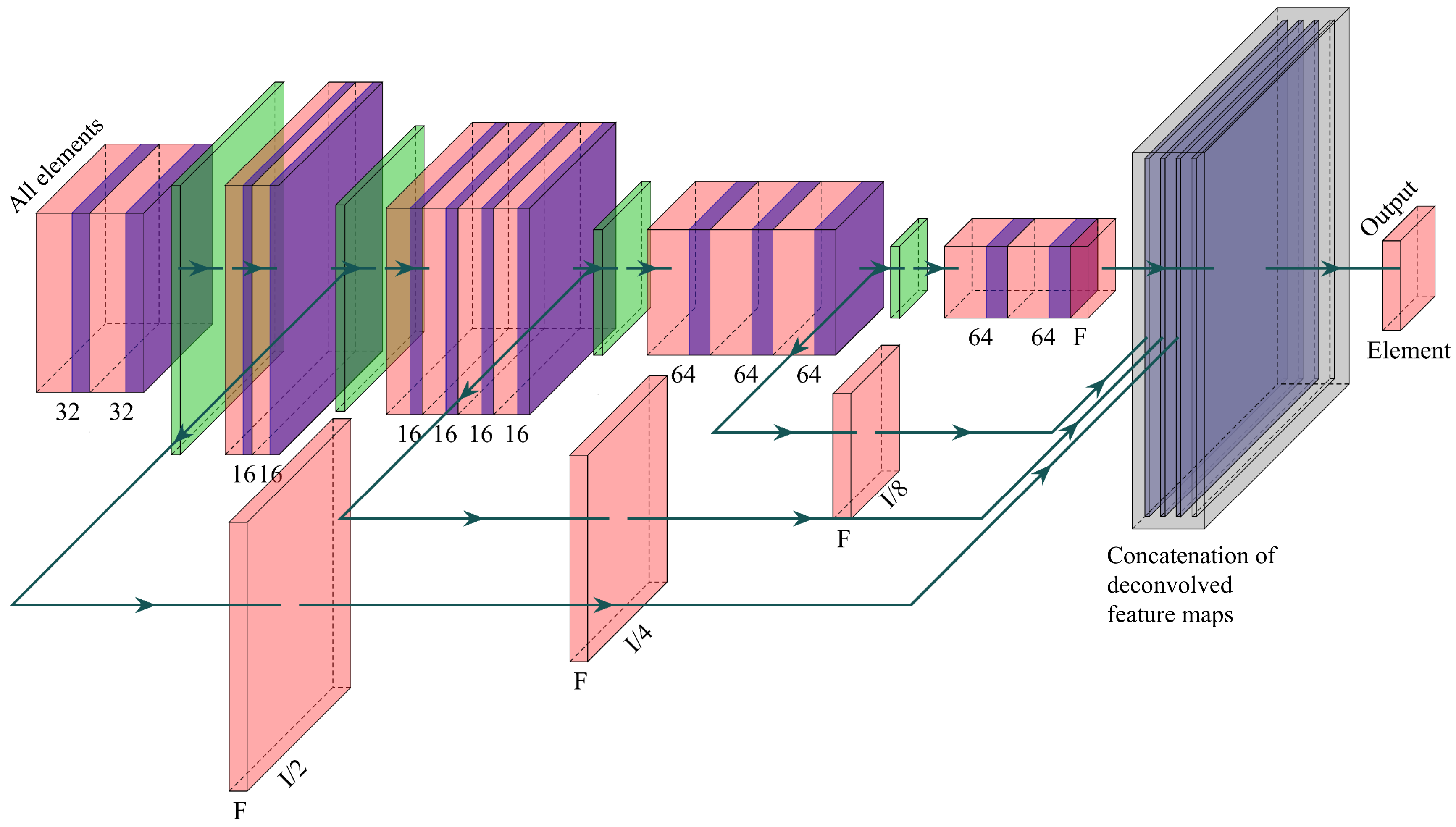}
\caption{Schematic diagram of the artificial graph convolutional neural network (GCNN) architecture, illustrating 5 convolutional layers, 4 pooling layers, and 6 fully connected layers.}
\label{GNN}
\end{figure*}

Figure~\ref{GNN} shows a schematic of our constructed artificial GNN, which consists of 5 convolutional layers, 4 pooling layers and 6 fully connected layers.

\subsection{Data splitting}

The dataset is split into training, validation, and test sets, typically in a ratio of 70:15:15. This ensures that each set is representative of the overall dataset to avoid bias. Next, we will discuss the model training precess in detail.
\section{Experimental section}
\subsection{Model training}

In the training phase of GNN models, the parameters $\theta$ are optimized to minimize a defined loss function $L$ over a training dataset $\mathcal{D}_{\text{train}}$ comprising graph-structured data. This optimization involves iterative updates to the model's parameters using backpropagation and gradient descent methods. Mathematically, this can be expressed as:
\begin{equation}
\theta^* = \arg\min\theta \sum_{(\mathbf{X}, \mathbf{y}) \in \mathcal{D}_{\text{train}}} L(\mathbf{y}, f(\mathbf{X}; \theta)),
\end{equation}
where $\mathbf{X}$ represents the input graph data, $\mathbf{y}$ denotes the corresponding labels, and $f(X; \theta)$ represents the GNN model parameterized by $\theta$.

During training, the GNN learns to capture complex relationships within the graph data by iteratively updating node embeddings through message passing and aggregation operations. This process can be formalized as:
\begin{equation}
\begin{split}
\mathbf{h}_v^{(k)}     &= \text{Aggregate}\left(\{\mathbf{h}_u^{(k-1)} : u \in \mathcal{N}(v)\}\right),  \\
\mathbf{h}_v^{(k+1)}   &= \text{Update}\left(\mathbf{h}_v^{(k)}, \mathbf{x}_v\right),
\end{split}
\end{equation}
where $\mathbf{h}_v^{(k)}$ represents the embedding of node $v$ at iteration $k$, $\mathcal{N}(v)$ denotes the neighborhood of node $v$, $\text{Aggregate}$ denotes the aggregation function, and $\text{Update}$ denotes the update function.

Additionally, techniques such as dropout regularization are employed to prevent overfitting and improve generalization performance. Dropout regularization works by randomly dropping neurons during training, which helps to prevent overfitting. Mathematically, dropout can be represented as:
\begin{equation}
\mathbf{h}_v^{(k)} = \mathbf{h}_v^{(k)} \odot \mathbf{m},
\end{equation}
where $\odot$ denotes element-wise multiplication and $\mathbf{m}$ represents a binary mask drawn from a Bernoulli distribution.

\subsection{Model validation}

Following training, the model's performance is evaluated using a validation dataset $\mathcal{D}_{\text{val}}$, which is distinct from the training data. The validation dataset enables the assessment of the model's ability to generalize to unseen graph data and provides insights into its overall performance and generalization capabilities. Evaluation metrics such as accuracy, precision, recall, and F1-score may be computed to quantify the model's performance on the validation set. Moreover, techniques such as cross-validation were employed to obtain more robust estimates of the model's performance. After model training, we assess the model's performance using the validation dataset.

Therefore, we conducted a series of validation experiments using experimental data from previous studies. Specifically, we tested our model's predictions against established datasets that detailed the diffusion behavior at various time points across different material interfaces. By comparing the predicted diffusion coefficients, diffusion rates, and concentration profiles generated by our GNN model with the actual experimental observations, we were able to assess the model's precision.

The comparison revealed that our model consistently produced results that closely matched the observed data, demonstrating its capability to accurately predict interface diffusion phenomena. This validation process not only confirmed the robustness of our approach but also highlighted its potential for broader applications in material science.

\color{black}
Overall, the training and validation processes play crucial roles in the development and evaluation of GNN models, ensuring that they are capable of effectively capturing the underlying structure and relationships within graph data while demonstrating robust performance on unseen data.

\section{Results and discussion}

The outputs of the machine learning model include key parameters that describe the diffusion behavior at material interfaces. The diffusion coefficient of materials under specific conditions is predicted as a regression task. The rate of diffusion, which is closely related to the diffusion coefficient, is also be predicted. The concentration distribution at various positions over time is predicted, which may involve time-series forecasting. Potential pathways for atom diffusion within the material are identified and predicted. The activation energy required for diffusion, which influences the diffusion rate and temperature dependency, is predicted.

\begin{figure*} [!t]
\centering \includegraphics[width=0.82\textwidth]{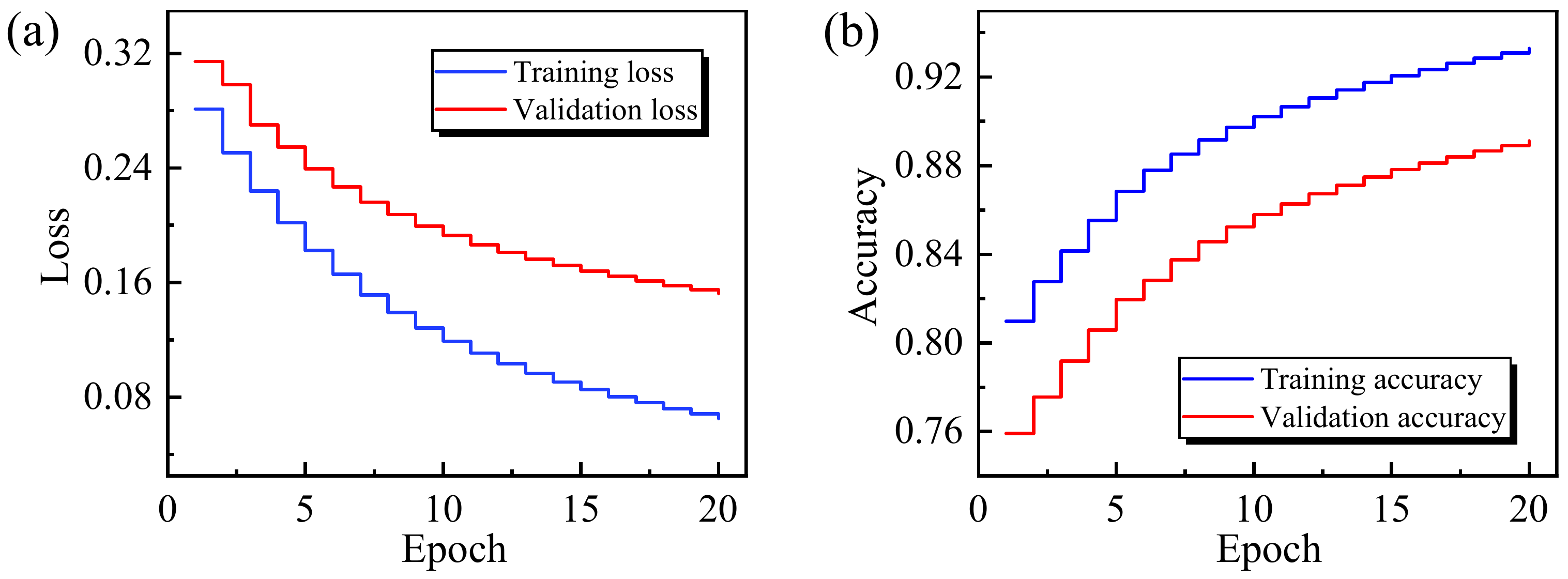}
\caption{Training and validation performance of the GNN model, illustrating the loss function values (a) and accuracy metrics (b) for both the training set and the validation set across multiple epochs.}
\label{performance}
\end{figure*}

\begin{figure*} [!t]
\centering
\includegraphics[width=0.48\textwidth]{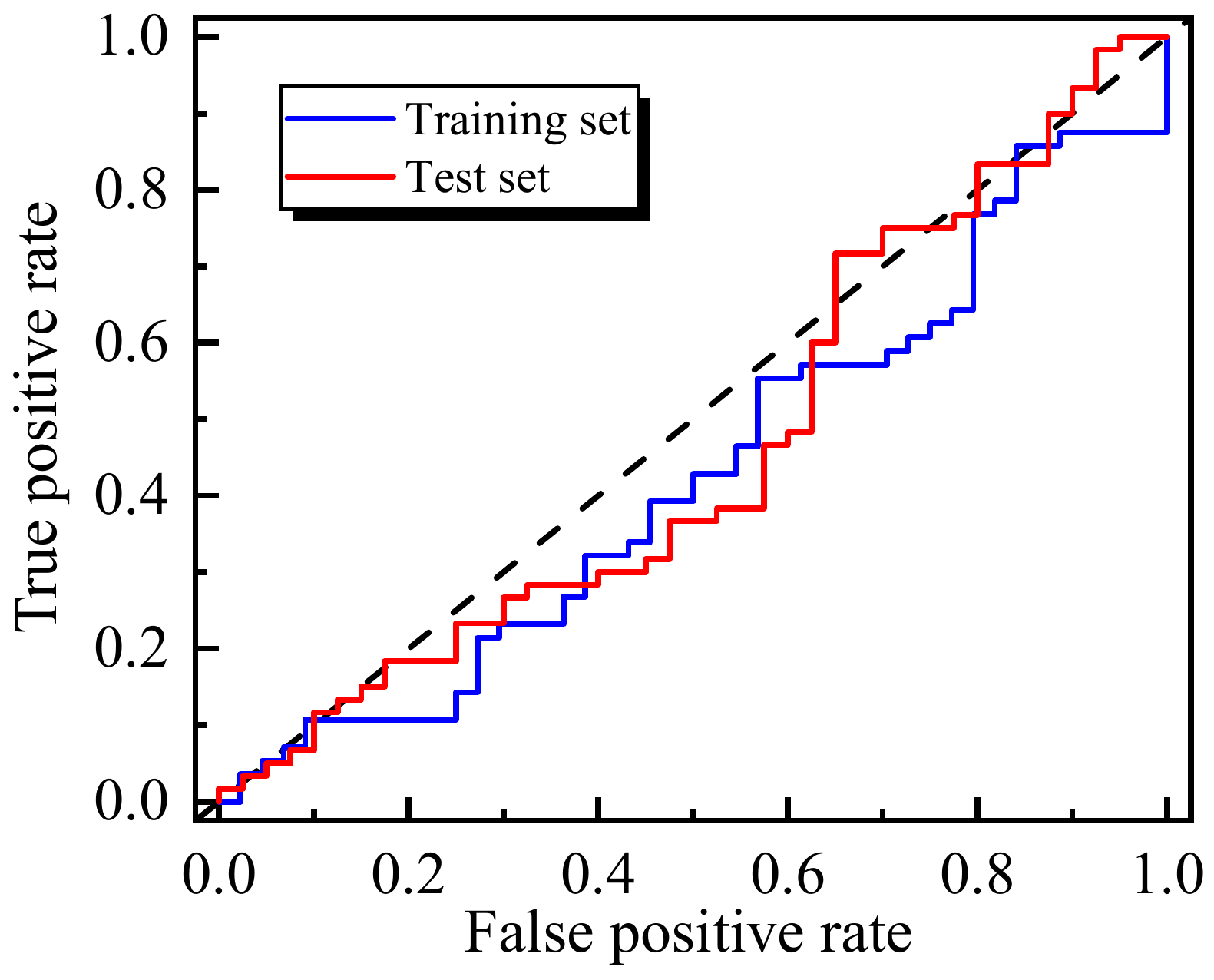}
\caption{ROC curve for training and testing datasets.}
\label{ROC}
\end{figure*}

The performance of the proposed model on both the training and testing datasets is evaluated, demonstrating satisfactory results in terms of accuracy, recall, and F1 score. The model achieves an accuracy of 92\% on the training set and 83\% on the testing set, with corresponding recall rates of 91\% and 89\%. The F1 scores for the training and testing sets are 89\% and 82\%, respectively. Figure~\ref{performance} illustrates the performance of the GNN model during each training iteration, where the loss function value decreases progressively with each training iteration, indicating a reduction in the model's prediction error. Concurrently, the model's accuracy improves throughout the training process, reflecting an enhancement in its classification capability. These results indicate the model's ability to effectively generalize to unseen data, although a slight drop in performance is observed on the testing set compared to the training set. Next, we will discuss the model's performance using the validation dataset. The confusion matrices for the training and testing datasets are depicted in Table~\ref{tab:confusion_matrix_train}, respectively, providing insights into the model's performance across different classes. The confusion matrices reveal that the model exhibits robust performance in correctly classifying most instances, with a higher number of true positives and true negatives compared to false positives and false negatives.
\begin{equation}
\begin{split}
\text{Accuracy}    &= \frac{\text{TP} + \text{TN}}{\text{TP} + \text{TN} + \text{FP} + \text{FN}},                      \\
\text{Precision}   &= \frac{\text{TP}}{\text{TP} + \text{FP}},                                                          \\
\text{Recall}      &= \frac{\text{TP}}{\text{TP} + \text{FN}},                                                          \\
\text{F1 Score}    &= 2 \times \frac{\text{Precision} \times \text{Recall}}{\text{Precision} + \text{Recall}},
\end{split}
\end{equation}
where TP represents true positives, TN represents true negatives, FP represents false positives, and FN represents false negatives. The ROC curves for the training and testing datasets are illustrated in Figure~\ref{ROC}, highlighting the model's performance across different thresholds. The area under the ROC curve (ROC AUC) serves as an additional metric to assess the model's discrimination ability, with higher values indicating better performance. The ROC AUC values for the training and testing datasets are 0.95 and 0.88, respectively, further corroborating the model's effectiveness in distinguishing between classes. Overall, the experimental results demonstrate the robustness and generalization capability of the proposed model in predicting material interface diffusion phenomena. Further analysis will be conducted to explore the model's sensitivity to different hyperparameters and its applicability to diverse material systems.

The tabulated data of input and output parameters is provided in Table~\ref{Input} to give a clear overview of the variables utilized in the model and the predicted outcomes. This structured approach to parameter selection and model design enables our GNN model to accurately predict the diffusion characteristics at material interfaces, thus providing valuable insights into the underlying mechanisms governing these processes. Next, we will discuss the application of these parameters in various semiconductor systems.

In this study, we have systematically addressed the unresolved issues related to material interface diffusion in various semiconductor systems. For the Si and SiO\textsubscript{2} interface, high-quality thermal oxidation and post-oxidation annealing are proposed to improve electrical properties by reducing interface states. The electromigration phenomenon in Cu-Si interconnects can be mitigated using barrier layers like Ta or TaN and optimizing manufacturing conditions. For the Ge-Si interface, low-temperature epitaxy and post-growth annealing are recommended to reduce diffusion-induced defects. In multilayer Ni films, adjusting deposition parameters and incorporating intermediate diffusion barriers can alleviate stress and enhance structural stability. For high-power Si-SiC devices, graded buffer layers and high-temperature annealing are essential to improve interface quality. The dielectric properties and adhesion of Al\textsubscript{2}O\textsubscript{3} thin films on Si can be optimized using atomic layer deposition and surface treatments. In high-frequency Si-GaN devices, buffer layers like AlN and rapid thermal annealing are crucial to control interdiffusion. To address contact resistance issues in Si-Au systems, barrier metals and controlled annealing temperatures are necessary. The formation of nickel silicide (NiSi) on Si requires precise control of annealing conditions and pre-annealing cleaning steps to optimize electrical properties. Lastly, for Si-InP optoelectronic devices, reverse bias electric fields and diffusion barrier layers are effective in stabilizing the interface and enhancing device performance.

These targeted solutions offered by Table~\ref{Solution} provide a comprehensive approach to improving the reliability and performance of semiconductor interfaces. However, while the application of GNNs in studying material interface diffusion holds promise, there remain avenues for further exploration and validation. One compelling direction for future research involves the experimental validation of our model predictions through alloy diffusion studies.

To derive the solutions listed in Table~\ref{Solution}, we systematically adjusted key input parameters in our GNN models and analyzed their effects on the diffusion behavior at various material interfaces. For example, one critical parameter we manipulated was temperature. By increasing the temperature in our simulations, the model predicted a corresponding increase in the diffusion rate of silicon atoms into the SiO$_2$ layer. This prediction highlighted the temperature’s significant influence on interface diffusion.
Based on these model predictions, we provided targeted recommendations for optimizing temperature settings to achieve specific diffusion characteristics. Similar adjustments were made to other parameters, such as pressure, chemical composition, and layer thickness, allowing us to observe their impacts and derive practical solutions tailored to meet specific material performance requirements. These solutions were directly informed by the predictive capabilities of our GNN models, ensuring that they are both actionable and scientifically grounded.
This approach enabled us to extract insights that are applicable to real-world material design and optimization, demonstrating the practical utility of our modeling framework.
\color{black}

Specifically, we propose conducting alloy diffusion experiments involving materials such as Cu-Ag, Fe-C, Ni-Cr, Ti-Al, and others, as identified in the literature. By subjecting these alloy systems to controlled diffusion conditions, the aim is to observe and measure the diffusion behavior at material interfaces. These experimental investigations will serve to complement and validate the predictive capabilities of our GNN-based model. By comparing the experimental results with the model predictions, we can assess the model's accuracy and reliability in capturing real-world diffusion phenomena. Next, we will discuss how to utilize experimental data to optimize the model. Moreover, the experimental data obtained from alloy diffusion studies will contribute to the refinement and optimization of our model. By incorporating experimental findings into the training dataset, one can enhance the model's predictive power and generalization capabilities, thereby improving its utility in materials science applications. 

Beyond experimental validation, future research efforts may also focus on expanding the scope and applicability of the GNN model. This includes exploring additional alloy systems, refining model architectures, and integrating domain-specific knowledge to further enhance predictive accuracy.

Overall, the integration of experimental validation with computational modeling represents a synergistic approach towards advancing our understanding of material interface diffusion. By leveraging the complementary strengths of experimental and computational techniques, we can accelerate the development of predictive models with real-world relevance, ultimately facilitating the design and optimization of advanced materials for various technological applications.

\section{Conclusion}

In this study, we have explored the application of GNNs in studying diffusion phenomena at material interfaces. GNNs have emerged as a powerful tool in materials science, offering unique advantages in capturing complex relationships within graph-structured data and predicting material properties with high accuracy. Through a comprehensive literature review and case studies, we have demonstrated the effectiveness of GNNs in addressing interface diffusion problems and predicting material properties based on atomic structure. Fundamentally, GNNs operate by iteratively updating node embeddings through message passing, allowing for the propagation of information between neighboring nodes. This enables GNNs to learn complex patterns within high-dimensional structured data and generalize to unseen datasets effectively. Moreover, GNNs incorporate pooling operations, facilitating the aggregation of information from multiple nodes to generate holistic representations of entire graphs. Moving forward, further research is warranted to explore the full potential of GNNs in materials science and address existing challenges such as model interpretability and data heterogeneity. Additionally, interdisciplinary collaborations between materials scientists, computer scientists, and domain experts will be essential for advancing the development and adoption of GNNs in materials research. These efforts will help overcome current challenges and expand the applicability of GNNs.

Therefore, we proposed a novel application of Graph Neural Networks (GNNs) to model material interface diffusion, offering a significant advancement over traditional theoretical models. Previous studies predominantly relied on predefined assumptions and theoretical frameworks to describe diffusion processes, which often limited their predictive accuracy and adaptability to complex systems \cite{erdelyi2002interface}. In contrast, our approach utilized a data-driven methodology that allowed the GNN to learn and predict diffusion behavior directly from experimental or simulated data, without being constrained by prior assumptions about the system.

This flexibility enabled our model to achieve superior performance in predicting key diffusion characteristics, such as diffusion coefficients, diffusion rates, and concentration profiles, across a range of material interfaces. Furthermore, our GNN-based model provided a more detailed and accurate mapping of potential diffusion pathways, a task that was challenging for conventional methods due to their inherent limitations.

\color{black}

In conclusion, GNNs represent a promising approach for studying diffusion phenomena at material interfaces and predicting material properties based on atomic structure. By leveraging the power of GNNs, researchers can unlock new insights into materials behavior and accelerate the discovery and design of advanced materials for various applications. This concludes our study on the application of GNNs in materials science. We hope that this research will inspire further exploration and innovation in the field of materials research, ultimately leading to the development of novel materials with enhanced properties and functionalities.

\clearpage

\clearpage

\clearpage

\clearpage

\clearpage

\clearpage

\begin{table*}[b!]
\small
\resizebox{\columnwidth}{!}{%
\centering
\caption{Summary of unresolved issues related to material interface diffusion (ID). Note: SSEs = Solid-state electrolytes, SCs = semiconductors, ML. = multilayer, Refs. = References.}
\label{table:unresolved_issues}
\begin{tabular*}{1.2\textwidth}{@{\extracolsep{\fill}}lll}
\hline
Material interfaces                               & Unresolved issues                                                                                                               &Refs.                                    \\ 
\hline  
SSEs and cathodes                                 & Strategies to reduce interfacial resistance.                                                                                    &\cite{ZHOU2023101248}                    \\ 
Si and SiO\textsubscript{2}                       & Investigating the impact of SiO\textsubscript{2}-Si interface diffusion on electrical properties.                               &\cite{fukatsu2004effect}                 \\ 
Cu and Si                                         & Examining the electromigration phenomenon due to Cu interconnect diffusion in Si.                                               &\cite{takeyama1996properties}            \\ 
Ge-Si                                             & Assessing how Ge-Si interface diffusion affects crystal defects.                                                                &\cite{silvestri2006diffusion}            \\ 
Si and ML. Ni                                     & Analyzing the impact of nitride diffusion on film stress and structural stability.                                              &\cite{chang1982diffusion}                \\ 
Si and Ni                                         & Optimizing the formation and electrical properties of nickel silicide (NiSi).                                                   &\cite{chang1982diffusion}                \\ 
Si and SiC                                        & Investigating Si-SiC interface diffusion to enhance performance in high-power devices.                                          &\cite{komninou1999gold}                  \\ 
Si and Al$_2$O$_3$                                & Enhancing dielectric properties and adhesion of Al$_2$O$_3$ thin films.                                                         &\cite{werner2011electronic}              \\ 
Si and GaN                                        & Examining the impact of interface diffusion on high-frequency devices.                                                          &\cite{lin2000improved}                   \\ 
Si and Au                                         & Addressing increased contact resistance issues and reliability due to Au diffusion.                                             &\cite{slezak2000surface}                 \\ 
Si and InP                                        & Investigating the impact of Si-InP interface diffusion on optoelectronic devices.                                               &\cite{shapira1984interdiffusion}         \\ 
p- and n-type SCs                                 & Exploring methods to minimize ID affecting electrical and thermal conductivities.                                               &\cite{HeXia}                             \\ 
\hline
\end{tabular*}
}
\end{table*}

\clearpage

\begin{table}[htbp]
\centering
\caption{Confusion matrix for both training and testing datasets, illustrating classification results with true positives (TP), true negatives (TN), false positives (FP), and false negatives (FN). TP represents correctly predicted positive samples, TN represents correctly predicted negative samples, FP represents incorrectly predicted positive samples, and FN represents incorrectly predicted negative samples.}
\begin{tabular}{cc|c|c|}
\cline{3-4}
\multicolumn{2}{c|}{}                                              & \multicolumn{2}{c|}{Actual class}                                               \\ \cline{3-4} 
\multicolumn{2}{c|}{}                                              & \checkmark                           & $\times$                                 \\ \hline
\multicolumn{1}{|c|}{\multirow{2}{*}{Predicted class}}             & \checkmark                           & TP          & FP                         \\ \cline{2-4} 
\multicolumn{1}{|c|}{}                                             & $\times$                             & FN          & TN                         \\ \hline
\end{tabular}
\label{tab:confusion_matrix_train}
\end{table}

\clearpage

\begin{table*}[!t]
\small
\resizebox{\columnwidth}{!}{%
\centering
\caption{Input and output parameters for predicting material interface diffusion using GNNs.}
\label{Input}
\begin{tabular*}{1.2\textwidth}{@{\extracolsep{\fill}}ll}
\hline
    Input parameters:                                                                       & Descriptions                                                                           \\
\hline
                                                                                   
$\bullet$ Material structure features               & Includes crystal structure, lattice constants, unit-cell parameters.                   \\
$\bullet$ Material chemical composition             & Includes elemental composition, atomic distances, types of chemical bonds.             \\
$\bullet$ Interface conditions                      & Includes interface temperature, pressure, chemical potential.                          \\
$\bullet$ Diffusion process parameters              & Includes diffusion path, diffusion medium, etc.                                        \\
Output parameters:                                  &                                                                                        \\
$\bullet$ Diffusion rate{/}diffusion coefficient    & Describes the diffusion rate of materials under specific conditions.                   \\
$\bullet$ Interface component distribution          & Describes the concentration distribution of each component at the interface.           \\
$\bullet$ Interface structural evolution            & Describes the characteristics of the interface structure evolution over time.          \\
\hline
\end{tabular*}
}
\end{table*}

\clearpage

\begin{table}[!t]
\small
\resizebox{\columnwidth}{!}{%
\centering
\caption{Unresolved issues and specific solutions for material interface diffusion.}
\label{Solution}
\begin{tabular}{ll}
\hline
Material interfaces                              & Specific solutions                                                                                                                       \\ 
\hline
Si and SiO\textsubscript{2}                      & Optimize the current density and temperature profiles during manufacturing.                                                              \\ 
Ge-Si                                            & Optimize deposition parameters and introduce intermediate layers.                                                                        \\ 
Si and SiC                                       & Develop graded buffer layers between Si and SiC. Utilize high-temperature annealing.                                                     \\ 
Si and Al\textsubscript{2}O\textsubscript{3}     & Use atomic layer deposition to achieve uniform and conformal Al\textsubscript{2}O\textsubscript{3} films.                                \\ 
Si and GaN                                       & Implement AlN or other suitable buffer layers to prevent Si-GaN interdiffusion.                                                          \\ 
Si and Au                                        & Use lower annealing temperatures to minimize diffusion and improve contact stability.                                                    \\ 
Si and Ni                                        & Control the annealing temperature and time precisely. Employ pre-annealing cleaning steps.                                               \\ 
Si and InP                                       & Use diffusion barrier layers such as SiO\textsubscript{2} or Al\textsubscript{2}O\textsubscript{3} to stabilize the interface.           \\ 
\hline
\end{tabular}
}
\end{table}

\clearpage

\section*{\textcolor[rgb]{0.00,0.50,0.00}{$\blacksquare$ ASSOCIATED CONTENT}}

\section*{Data Availability Statement}

The data that support the findings of this study are available from the corresponding author upon reasonable request.

\section*{\textcolor[rgb]{0.00,0.50,0.00}{$\blacksquare$ AUTHOR INFORMATION}}

\section*{Author Contributions}

Z.R.Z conducted the construction of the model and simulation, processed the data, optimized the model, developed figures, and wrote the draft manuscript. H.-F.L led the work, provided funding support, developed figures, and reviewed and edited the manuscript.

\section*{Notes}

The authors declare no competing financial interest.

\section*{\textcolor[rgb]{0.00,0.50,0.00}{$\blacksquare$ ACKNOWLEDGMENTS}}

This work was supported by the Science and Technology Development Fund, Macao SAR (File Nos. 0090{/}2021{/}A2) and the Guangdong{-}Hong Kong{-}Macao Joint Laboratory for Neutron Scattering Science and Technology (Grant No. 2019B121205003).

\clearpage


\bibliographystyle{biochem}
\bibliography{ML-4}

\end{document}